\documentstyle[12pt]{article}
\input{psfig}
\makeatletter
\@addtoreset{equation}{section}
\makeatother

\begin{document}
\title{
 Gravitational Phase Transition of Fermionic Matter
in a General-Relativistic Framework
 }
\author{
Neven Bili\'c$^1$
and
Raoul D.~Viollier$^2$
 \\
 $^1$Rudjer Bo\v{s}kovi\'{c} Institute, 10000 Zagreb, Croatia \\
$^2$Department of Physics,
University of Cape Town, \\ Rondebosch 7701, South Africa
}
\date{\today}
\maketitle
\begin{abstract}
The Thomas-Fermi model  at finite temperature is extended
to describe a system of self-gravitating weakly interacting
massive fermions in a general-relativistic framework.
The existence and properties of the gravitational phase transition
in this model are investigated numerically. It is shown that,
by cooling a nondegenerate gas of weakly interacting massive
fermions below some critical temperature, a condensed phase emerges,
consisting of quasidegenerate fermion stars. For fermion masses
of 10 to 25 keV, these fermion stars may very well provide
an alternative explanation for the supermassive compact dark
objects that are observed at galactic centers.
\end{abstract}
%

\section{Introduction}
The ground state of a
 condensed cloud of fermionic matter,
 interacting only gravitationally
 and having a
 mass $M$ below the Oppenheimer-Volkoff (OV) limit \cite{opp},
 is a cold fermion star
 in which the degeneracy pressure
balances the gravitational attraction of the fermions.
Degenerate stars of fermions in the mass range between
  10  and  25 keV are particularly
interesting~\cite{vio}, as they could explain, without resorting
to the black-hole hypothesis, at least some of the features
 observed around the supermassive compact dark objects
 with masses in the range of
 $M=10^{6.5}$ to $10^{9.5}$ solar masses, that are
reported to exist at the centers of a number of
galaxies~\cite{Tonr,Dres10,Dres11,Korm12,Korm13,Korm14}, including
our own~\cite{Lacy15,Lacy16}, and quasistellar objects (QSO)
\cite{Bell,Zeld,Blan,Bege}.  Indeed, there is little difference
between a supermassive black hole and a fermion star of the same
mass near the OV limit, a few Schwarzschild radii away
from the object \cite{bil2,tsi}.

The purpose of
this paper is to study,
in the framework
of a general-relativistic Thomas-Fermi model,
 the formation of such a star
 that could have taken place
 in the early universe
 shortly after the nonrelativistic
 fermionic matter began to dominate the radiation.
 This system was previously studied in the
 Newtonian approximation~\cite{thi,her1,her2,bau,mes,bil1},
 and it was shown that the
  canonical and grand-canonical ensembles
     for such a system
have a nontrivial thermodynamical limit~\cite{thi,her1,her2}.
Under certain conditions these systems
 will undergo a phase transition
 that is accompanied by a gravitational collapse~\cite{mes,bil1}.
 The phase transition
occurs uniquely
in the case of
 the attractive gravitational interaction of
neutral fermions.
As the phase transition does not happen for particles
obeying Bose-Einstein or Boltzmann statistics,
this phenomenon is quite distinct from the usual gravitational
clustering of collisionless dark-matter particles.
Gravitational
condensation will also take place if the  fermions have an
additional short-range weak interaction, as neutrinos, neutralinos,
gravitinos,
 and other
weakly interacting massive particles do.

 Effects of general relativity
 cannot be neglected when
 the total mass of the system is close to the
 OV limit~\cite{opp}.
 There are three main features that distinguish the
 general-relativistic
 Thomas-Fermi model
  from the Newtonian one:
  {\it i}) the equation of state is relativistic,
  {\it ii}) the temperature and chemical potential are
  metric-dependent local quantities, and
  {\it iii}) the gravitational potential satisfies Einstein's field
  equations instead of Poisson's equation.

This paper is organized as follows:
In Section~\ref{sec2} we briefly discuss
the nonrelativistic Thomas-Fermi model
at finite temperature.
In Section~\ref{sec3}  this model is extended within
a general-relativistic framework.
 In Section~\ref{sec4} we discuss the
solution at zero and finite
 temperature  and, in particular,
the conditions under which
the first-order gravitational phase transition occurs.
Conclusions are drawn
in Section~\ref{sec6}
and, finally, in Appendix~\ref{app1}, we prove a theorem on
the extremal properties of the free energy.

\section{Thomas-Fermi model in Newtonian gravity}
\label{sec2}

Consider a system of $N$ fermions
of mass $m$
interacting only gravitationally, confined in
a spherical cavity of radius $R$, in equilibrium
at a finite temperature $T$.
For large $N$, we can
assume that the fermions move in a spherically
 symmetric
 mean-field potential
$V(r)$ which satisfies Poisson's equation
\begin{equation}\label{eq00}
\frac{1}{r}\frac{d^2}{dr^2}(rV) = 4\pi G  m^2 n,
\end{equation}
$G$ being the gravitational constant.
The number density  of the
fermions (including antifermions) $n$ can be expressed
in terms of the Fermi-Dirac distribution
(in units $\hbar=c=k=1$)
\begin{equation}\label{eq02}
n(r) = g \int \frac{d^3q}{(2\pi)^3}
\left[1+\exp \left(\frac{q^2}{2mT}+\frac{V(r)}{T}-\frac{\mu}{T}
\right)\right]^{-1}.
\end{equation}
Here
$g$ denotes the combined spin-degeneracy factors of
the neutral
fermions and antifermions, i.e., $g$ is 2 or 4 for Majorana
 or Dirac fermions, respectively.
 For each solution
 $V(r)$ of Eq.~(\ref{eq00}),
the chemical potential $\mu$ is adjusted so that
 the constraint
\begin{equation}\label{eq04}
\int d^3r n(r) = N
\end{equation}
is satisfied.
It may be shown that a particular spherically symmetric
configuration $\bar{n}(r)$ will satisfy
Eqs.~(\ref{eq00})-(\ref{eq04})
if and only if it extremizes the free
 energy functional
  defined as~\cite{her1}
\begin{eqnarray}
F[n]
\! & \! = \! & \!
     \mu[n] N     -
 \frac{1}{2}
\int d^3r n(r) V[n]
\nonumber  \\
     &  &
- Tg  \int\frac{d^3rd^3q}{(2\pi)^3}
\ln\left( 1+\exp \left(-\frac{q^2}{2mT}-
\frac{V[n]}{T}+\frac{\mu[n]}{T}\right)\right),
\label{eq06}
\end{eqnarray}
where $V[n]$ and $\mu[n]$ are
implicit functionals of $n(r)$ through
Eqs.~(\ref{eq00})-(\ref{eq04}).
For a physical
solution, we have to require that the free energy is
minimal, i.e.,
\begin{equation}\label{eq08}
\left. \frac{\delta F}{\delta n}\right|_{\bar{n}} = 0,
\;\;\;\;
\left. \frac{\delta^2 F}{\delta n^2}\right|_{\bar{n}} \geq 0 .
\end{equation}
The set of self-consistency equations
 (\ref{eq00})-(\ref{eq04}), together with
 (\ref{eq08}), comprises the
 nonrelativistic
 gravitational Thomas-Fermi model.

It may be easily shown that
the following scaling property holds:
  If the potential energy
 $V(r)$ is a solution to the self-consistency
 equations (\ref{eq00})-(\ref{eq04}),
  then the rescaled
 $\tilde{V}=A^4 V(Ar)$,
 with $A>0$,
 is also a solution with
 the rescaled
 fermion number
 $\tilde{N}=A^{3}N$,
 radius
 $\tilde{R}=A^{-1}R$,
 and temperature
 $\tilde{T}=A^4T$.
 This property, which will be  referred to as
 {\em nonrelativistic scaling},
 implies the existence of
 a thermodynamic limit
 of
 $N^{-7/3}F$,
 with
 $N^{1/3}R$ and
 $N^{-4/3}T$
 approaching constant values
  for
 $N\rightarrow\infty$.
 In this limit,
 the Thomas-Fermi equation becomes exact
 \cite{her1,her2}.

\section{Thomas-Fermi model in general relativity}
\label{sec3}

Consider a self-gravitating
gas consisting of $N$ fermions of mass
$m$ in equilibrium within a sphere of radius $R$.
We denote by
$p$, $\rho$, $n$, and $\sigma$ the  pressure,
energy density, particle number density, and
entropy density of the gas, respectively.
The metric generated by the mass distribution
is  static, spherically symmetric, and asymptotically
flat, i.e.,
\begin{equation}
ds^2=\xi^2 dt^2 -(1-2{\cal{M}}/r)^{-1} dr^2 -
     r^2(d\theta^2+\sin \theta d\phi^2).
\label{eq13}
\end{equation}
 Einstein's field equations are then given by
\begin{equation}
\frac{d\xi}{dr}=\xi\frac{{\cal M}+4\pi r^3 p}{r(r-2{\cal{M}})} \, ,
\label{eq40}
\end{equation}
\begin{equation}
\frac{d{\cal{M}}}{dr}=4\pi r^2 \rho,
\label{eq42}
\end{equation}
with the boundary conditions
\begin{equation}
\xi(R)=\left(1-\frac{2{\cal{M}}(R)}{R}\right)^{1/2}
\, ; \;\;\;\;\;
{\cal{M}}(0)=0.
\label{eq44}
\end{equation}
The  equation of state may be represented
in a parametric form \cite{ehl}
\begin{equation}
n = g \int \frac{d^3q}{(2\pi)^3}\,
\frac{1}{1+e^{E/\bar{T}-\bar{\mu}/\bar{T}}} \, ,
\label{eq30}
\end{equation}
\begin{equation}
\rho = g \int \frac{d^3q}{(2\pi)^3}\,
\frac{E}{1+e^{E/\bar{T}-\bar{\mu}/\bar{T}}} \, ,
\label{eq31}
\end{equation}
\begin{equation}
p = g \bar{T} \int \frac{d^3q}{(2\pi)^3}\,
\ln (1+e^{-E/\bar{T}+\bar{\mu}/\bar{T}}) \, ,
\label{eq32}
\end{equation}
where
$g$ denotes the spin degeneracy factor
 and $E=\sqrt{m^2+q^2}$.
 The quantities
$\bar{T}$ and $\bar{\mu}$ are
the local temperature and chemical potential,
respectively.
As discussed in Appendix \ref{app1},
thermodynamic and hydrostatic equilibrium
in the presence of gravity
implies
\begin{equation}
\bar{T}(r) =\frac{T}{\xi(r)}\, ; \;\;\;\;\;\;
\bar{\mu}(r)=\frac{\mu}{\xi(r)} \, .
\label{eq24a}
\end{equation}
The constants
$T$ and $\mu$  are the
temperature and chemical potential at infinity.
Although the matter is absent at $r=\infty$,
the temperature at infinity has
  a physical
meaning: T is the "red shifted" temperature
\cite{bar}
 of the black-body radiation of a gravitating object
 in equilibrium at finite temperature
measured at infinity.
In other words, our gravitating object is in equilibrium
with a heat bath which could be thought of as
 a black-body radiation in the empty space
surrounding the object.
As a consequence of Eq. (\ref{eq24a}),
different gravitating
configurations of the same size ``in contact" with the same
heat bath may have different surface temperatures.
Therefore, the relevant equilibrium parameter is
the temperature of the heat bath, $T$, and
not the surface temperature of the gravitating object.
 Particle number conservation  yields the constraint
\begin{equation}
\int_0^Rdr\, 4\pi r^2 (1-2{\cal{M}}/r)^{-1/2}\, n(r)=N .
\label{eq45}
\end{equation}
 Given the temperature at infinity  $T$,
the set of self-consistency
equations (\ref{eq40})-(\ref{eq45}) defines
the  general-relativistic Thomas-Fermi equation.
One additional important requirement is that a solution
to the equations (\ref{eq40})-(\ref{eq45})
should  minimize
the free energy.
Based on the considerations given in Appendix \ref{app1},
the free energy may be written
in the form
\begin{equation}
F =M+\mu N -\int dr4\pi r^2 \xi
 (1-2{\cal{M}}/r)^{-1/2}
(p+\rho),
\label{eq27}
\end{equation}
with  $M={\cal M}(R)$.
The theorem proved in Appendix A guarantees that
 solutions to Eqs. (\ref{eq40})-(\ref{eq45})
 extremize the free energy $F$,
i.e., the free-energy functional assumes either
a maximum or a minimum.
We only have to find out which of the solutions are
maxima and discard them as unphysical.

Next we  show  that, in the
Newtonian limit, we recover the
usual Thomas-Fermi model
as defined in Section~\ref{sec2}.
Introducing
the nonrelativistic chemical potential
$\mu_{NR}=\mu-m$
and
 the approximations
$\xi=e^{\varphi}\simeq 1+\varphi$,
$E\simeq m+q^2/2m$
and
  ${\cal M}/r \ll 1$ ,
 we arrive at the Thomas-Fermi self-consistency
 equations \cite{mes,bil1} in the form
\begin{equation}
n=\frac{\rho}{m}
 = g \int \frac{d^3q}{(2\pi)^3}\,
\left(1+\exp(\frac{q^2}{2mT}+\frac{m}{T}\varphi
 -\frac{\mu_{NR}}{T}) \right)^{-1} \, ,
\label{eq46}
\end{equation}
\begin{equation}
\frac{d\varphi}{dr}=\frac{{\cal M}}{r^2} \, ;
\;\;\;\;
\frac{d{\cal{M}}}{dr}=4\pi r^2 \rho \, ,
\label{eq47}
\end{equation}
\begin{equation}
\varphi(R)=-\frac{m N}{R}
\, ; \;\;\;
{\cal{M}}(0)=0,
\label{eq41}
\end{equation}
\begin{equation}
\int_0^R dr\,4\pi r^2 n(r)=N,
\end{equation}
which are equivalent to
 the set of equations (\ref{eq00})-(\ref{eq04}).
The free energy  (\ref{eq27}) in the Newtonian limit yields
\begin{equation}
F=m N +\mu_{NR} N - \frac{1}{2}\int_0^R dr \, 4\pi r^2 n\varphi
-\int_0^R dr \, 4\pi r^2 p \, ,
\label{eq43}
\end{equation}
with
\begin{equation}
 p= g T\int \frac{d^3q}{(2\pi)^3}\,
\ln\left(1+\exp(-\frac{q^2}{2mT}-\frac{m}{T}\varphi
 +\frac{\mu_{NR}}{T}) \right) \, ,
\label{eq48}
\end{equation}
which, up to a constant, is equal the nonrelativistic Thomas-Fermi free
energy (\ref{eq06}).

A straightforward thermodynamic limit
$N\rightarrow\infty$,
as discussed by
Hertel, Thirring, and Narnhofer \cite{her1,her2},
is not directly applicable
 in the general-relativistic case.
 First,
 in contrast to the Newtonian case,
 there exists
  a limiting configuration
with maximal $M$ and $N$ (the OV limit)
at zero temperature,
 and, as we shall shortly demonstrate,
also at finite temperature.
Second,
the scaling property of the relativistic
Thomas-Fermi equation,
which will be referred to as
 {\em relativistic scaling},
is quite distinct from
 nonrelativistic scaling.
This scaling property may be formulated as follows:
  If the configuration
 $\{\xi(r),{\cal{M}}(r)\}$ is a solution to the self-consistency
 equations (\ref{eq40})-(\ref{eq45}), then the
 configuration
 $\{\tilde{\xi}=\xi(A^{-1}r),
 \tilde{\cal M}=A{\cal{M}}(A^{-1}r);A>0\}$
 is also a solution with
 the rescaled
 fermion number
 $\tilde{N}=A^{3/2}N$,
 radius
 $\tilde{R}=AR$,
 asymptotic temperature
 $\tilde{T}=A^{-1/2}T$,
 and fermion mass
 $\tilde{m}=A^{-1/2}m$.
 The free energy is then rescaled as
 $\tilde{F}=AF$.
 Hence, there exists a thermodynamic limit
 of
 $N^{-2/3}F$,
 with
 $N^{-2/3}R$,
 $N^{1/3}T$,
and  $N^{1/3}m$
 approaching constant values
  when
 $N\rightarrow\infty$.

\section{Numerical integration}
\label{sec4}
In the following we use the units in which
$G=1$.
We choose appropriate length, mass and fermion number scales
$a$, $b$, and $c$  respectively, such that
\begin{equation}
a=b=
\sqrt{\frac{2}{g}}\,
\frac{1}{m^2},
\;\;\;\;
c=
\frac{b}{m}  ,
\label{eq81}
\end{equation}
or,  restoring $\hbar$, $c$, and $G$,
we have
\begin{equation}
a=
\sqrt{\frac{2}{g}} \,
\frac{\hbar M_{\rm{Pl}}}{c m^2}
=1.3185 \times 10^{10}\,
\sqrt{\frac{2}{g}} \,
\left(
\frac{15{\rm keV}}{m}
\right)^2
{\rm km},
\end{equation}
\begin{equation}
b=
\sqrt{\frac{2}{g}} \,
\frac{M_{\rm{Pl}}^3}{m^2}
=0.8929 \times 10^{10}\,
\sqrt{\frac{2}{g}} \,
\left(
\frac{15{\rm keV}}{m}
\right)^2
M_{\odot}
\end{equation}
\begin{equation}
c=
\sqrt{\frac{2}{g}} \,
\frac{M_{\rm{Pl}}^3}{m^3}
=5.5942 \times 10^{71}\,
\sqrt{\frac{2}{g}} \,
\left(
\frac{15{\rm keV}}{m}
\right)^3
\label{eq82}
\end{equation}
where $M_{\rm{Pl}}=\sqrt{\hbar c/G}$
denotes the Planck  and $M_{\odot}$ the solar mass.

We are looking
 for a solution of the Thomas-Fermi problem
as a function of temperature.
For numerical convenience,
let us introduce a new parameter
\begin{equation}
\alpha=\frac{\mu}{T}
\label{eq86}
\end{equation}
and the substitution
\begin{equation}
\xi=\frac{\mu}{m}(\Phi+1)^{-1/2}.
\label{eq87}
\end{equation}
Using this and (\ref{eq81}),
Eqs.~(\ref{eq30})-(\ref{eq32}) may be written
in the form
\begin{equation}
n=\frac{1}{\pi^2}
               \int^{\infty}_{0} dy\,
\frac{y^2}{1+\exp [(\sqrt{(y^2+1)/(\Phi+1)}-1)\alpha]} \, ,
\label{eq83}
\end{equation}
\begin{equation}
\rho=\frac{1}{\pi^2}
               \int^{\infty}_{0}dy\,
\frac{y^2\sqrt{y^2+1}}
{1+\exp [(\sqrt{(y^2+1)/(\Phi+1)}-1)\alpha]} \, ,
\label{eq84}
\end{equation}
\begin{equation}
p=\frac{1}{3\pi^2}
               \int^{\infty}_{0}dy\,\frac{y^4
}{\sqrt{y^2+1}
(1+\exp [(\sqrt{(y^2+1)/(\Phi+1)}-1)\alpha]} \, .
\label{eq85}
\end{equation}
In this way, both the fermion mass and the
 chemical potential are eliminated from the
equation of state.

The field equations  (\ref{eq40}) and (\ref{eq42})   now read
\begin{equation}
\frac{d\Phi}{dr} =
-2(\Phi+1)\frac{{\cal M}+4\pi r^3 p}{r(r-2{\cal{M}})} \, ,
\label{eq88}
\end{equation}
\begin{equation}
\label{eq89}
\frac{d{\cal{M}}}{dr}=4\pi r^2 \rho.
\end{equation}
To these two equations we add
\begin{equation}
\frac{d{\cal N}}{dr}=4\pi r^2 (1-2{\cal{M}}/r)^{-1/2} n
\label{eq93}
\end{equation}
imposing
 the particle-number constraint as
a condition at the boundary
\begin{equation}
{\cal N}(R)=N.
\label{eq94}
\end{equation}
Equations (\ref{eq88})-(\ref{eq93})
should be integrated using
the boundary conditions at the origin
\begin{equation}
\Phi(0)=\Phi_0 > -1
\, ; \;\;\;\;\;
{\cal{M}}(0)=0;
\, ; \;\;\;\;\;
{\cal{N}}(0)=0.
\label{eq90}
\end{equation}
The parameter $\Phi_0$,
which is uniquely related to the central
density and pressure,
will eventually be fixed
by the requirement (\ref{eq94}).
For $r\geq R$, the function $\Phi$ yields
the usual empty-space Schwarzschild
solution
\begin{equation}
\Phi(r)=\frac{\mu^2}{m^2}
\left(1-\frac{2 M}{r}\right)^{-1}-1\, ,
\label{eq91}
\end{equation}
  with
\begin{equation}
M={\cal M}(R)=\int_0^R dr 4\pi r^2 \rho(r) .
\label{eq92}
\end{equation}

We now show that
a solution to
the general-relativistic Thomas-Fermi
equation exists
provided the number of fermions
is smaller than a
certain number $N_{\rm max}$
that depends on
$\alpha$ and $R$.
From  (\ref{eq84}) and
  (\ref{eq85})   it follows that,
for any $\alpha>0$,
the equation of state $\rho(p)$ is an
infinitely smooth function and
$d\rho /dp > 0$ for $p > 0$.
Then, as shown by
Rendall and Schmidt
\cite{ren},
 there exists for any value of
the central density $\rho_0$
a unique static, spherically symmetric solution of
the field equations
with $\rho \rightarrow 0$  as
$r$ tends to infinity.
In that limit
${\cal M}(r)\rightarrow\infty$
and
${\cal N}(r)\rightarrow\infty$,
as may be easily seen
by analyzing
the
$r\rightarrow \infty$
limit of
Eqs. (\ref{eq88}) and (\ref{eq89}).
However, the enclosed mass
$M$ and the number of fermions $N$
within a given radius $R$ will be finite.
We can then cut off the matter
from $R$ to infinity
and join the interior solution onto
the empty space Schwarzschild
exterior solution
by making use of
equation
 (\ref{eq91}).
 This equation together with
 (\ref{eq86})
 fixes the chemical potential
 and the temperature at infinity.
Furthermore, it may be shown that
our equation of state obeys
asymptotically  at high densities
a $\gamma$-law,
i.e.,
$\rho=$ const $n^{\gamma}$
and $p=(\gamma-1) \rho$,
with
$\gamma=4/3$.
In this case,
as is well known
\cite{har},
there exists a limiting configuration
$\{ \psi_{\infty}(r),{\cal{M}}(r)_{\infty}\}$
such that $M$ and $N$
approach nonzero values $M_{\infty}$
and $N_{\infty}$,
respectively,
as the central density
$\rho_{0}$
tends to infinity.
Thus, the quantity
 $N$ is a continuous function of
 $\rho_{0}$
on the interval $0 \leq \rho_0 < \infty$,
with $N=0$ for
 $\rho_{0}=0$,  and
 $N=N_{\infty}$
 as
 $\rho_{0}\rightarrow\infty$.
The range of $N$
depends on
$\alpha$ and $R$
and its
upper bound
 may be denoted by
$N_{\rm max}(R,\alpha)$.
Thus, for given
$\alpha$, $R$
and
$N<N_{\rm max}(R,\alpha)$
the set of self-consistency
equations (\ref{eq83})-(\ref{eq92}) has
at least one solution.

As is evident from the equation of state (\ref{eq30})-(\ref{eq32}),
if we do not fix the boundary and do not
constrain the particle number $N$, the
pressure (and the density) will never vanish
(except perhaps at $r=\infty$),
unless $T=0$. Thus, since we fix the boundary at $R$
and cut off the matter from $R$ to infinity,
the pressure (and the density) will have a discontinuity.
This characteristic of the non-relativistic Thomas-Fermi model
in atomic physics \cite{fay},
and Newtonian gravity \cite{her1,mes,bil1}
remains also in general relativity.
However, the density and the pressure decrease rapidly
with $r$, so if $R$ is chosen sufficiently
large, the pressure and the density at the boundary
will be extremely small.
Furthermore, the region $r>R$ is
never empty in reality, so that a positive pressure at the
boundary is more realistic than
a vanishing pressure.

The numerical procedure is now straightforward.
For a fixed, arbitrarily chosen
$\alpha$,
 we first  integrate
equations
 (\ref{eq88}) and (\ref{eq89})
 numerically
on the interval $(0,R)$ and find
solutions
 for various initial
$\Phi_0$.
Simultaneously integrating (\ref{eq93}),
we obtain
${\cal N}(R)$ as a function of $\Phi_0$.
The specific value of $\Phi_0$
is then determined
such that
${\cal N}(R)=N$.
The chemical potential $\mu$
corresponding to
this particular solution
 is given by
(\ref{eq91}).
If we now eliminate $\mu$
using (\ref{eq86}), we finally get the
parametric dependence on temperature
through $\alpha$.

\begin{figure}
\vbox{\vskip230pt
\includegraphics{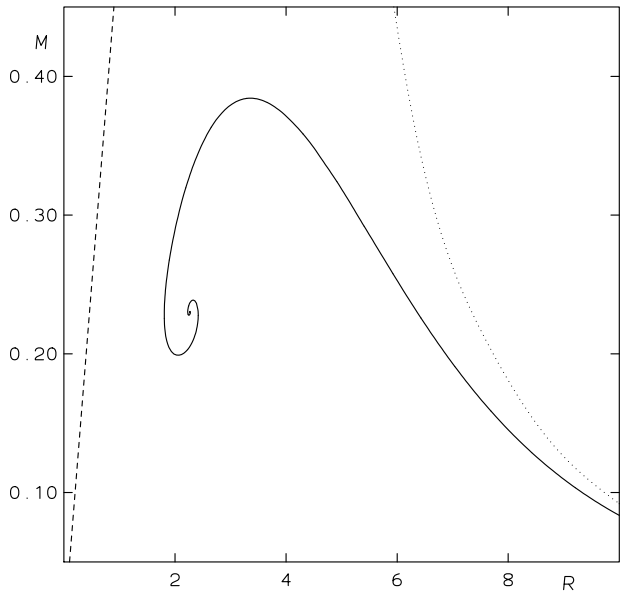}}
\caption{Mass versus radius for fermion stars at
zero temperature
in the general-relativistic framework
 (solid line)
 compared with the corresponding
Newtonian approximation (dotted line).
Dashed line is the black hole limit $M=R/2$.}
\label{fig1}
\end{figure}

Let us first discuss
a degenerate fermion gas ($T=0$)
 as a reference point
 that can also be compared with the
 well-known results by Oppenheimer and Volkoff
 \cite{opp}.
In this case,
the Fermi distribution in
(\ref{eq83})-(\ref{eq85})
becomes a step function that  yields
an elementary integral with the upper limit
$y_F=\sqrt{\Phi}$ related to the Fermi
momentum
\begin{equation}
q_F=m\sqrt{\Phi}.
\label{eq95}
\end{equation}
The equation of state can be expressed
in terms of elementary functions
of $\Phi$
\begin{equation}
n=\frac{1}{3\pi^2}\Phi\sqrt{\Phi},
\label{eq96}
\end{equation}
\begin{equation}
\rho=\frac{1}{8\pi^2} \left[(2\Phi+1)\sqrt{\Phi(\Phi+1)}
-{\rm Arsh}\sqrt{\Phi}\right] ,
\label{eq97}
\end{equation}
\begin{equation}
p=\frac{1}{24\pi^2} \left[(2\Phi-3)\sqrt{\Phi(\Phi+1)}
+3{\rm Arsh}\sqrt{\Phi}\right] .
\label{eq98}
\end{equation}
The radius of the star is naturally defined as
the point where the density vanishes.
At this point, owing to (\ref{eq96}),
$\Phi=0$.
Therefore, we integrate equations
(\ref{eq83})-(\ref{eq85})
starting from $r=0$ up to the point $R$
where $\Phi(R)=0$.
As a result, the quantities $M$, $N$, and $R$
are obtained
as functions of the parameter $\Phi_0$,
which is related to the central particle-number
  density through (\ref{eq96}).
In Fig.~\ref{fig1}, we plot the mass of the star as
a function of the radius $R$. The maximum of the curve
corresponds to the
OV limit \cite{opp}.
The limiting values are
$R_{\rm OV}=3.357$,
$M_{\rm OV}=0.38426$, and
$N_{\rm OV}=0.39853$,
in units of $a$, $b$, and $b/m$ respectively.
The curve
left from the maximum  represents unstable configurations
that curl up around the point corresponding to the
infinite central density limit.

We now  turn to the study of nonzero temperature.
The quantities
$T$, $N$, and, $R$ are free parameters in our model
and their
range and choice are dictated
by physics.
The temperature $T$ is restricted only to positive values.
The number of fermions $N$
 is restricted by the Oppenheimer-Volkoff limit.
 The radius
$R$ is theoretically unlimited; practically,
it should not exceed the order of interstellar distances.
It is known that a classical,
semidegenerate, isothermal configuration has no natural boundary
in contrast to the degenerate
 case of zero temperature,  where
for given $N$ (up to the Oppenheimer-Volkoff limit)
the radius $R$ is naturally fixed by the condition of
vanishing pressure and density.
At nonzero  temperature,
if we, e.g., fix only $N$ and $T$
and let $R\rightarrow \infty$, our gas will  occupy
the entire space, and hence $p$ and $\rho$ will
vanish everywhere.
If we do not restrict $N$ and integrate the
equations  on the interval $(0,\infty)$, $M$ and $N$ will
diverge at $\infty$.
In such a case one has to introduce a cutoff.
In the isothermal model of a similar kind
by Chau, Lake, and Stone \cite{cha},
a cutoff was chosen
at the radius $R$, where the energy density
was about six orders smaller than the central value.
Our choice is based on the following
considerations:
As in the Newtonian Thomas-Fermi model
\cite{bil1},
we expect that,
for given
number of fermions
 $N <  N_{\rm OV}$,
there exists a unique
configuration
that is  a solution
to the self-consistency equations
(\ref{eq30}) to (\ref{eq45})
and which
 becomes
a degenerate Fermi gas
at $T=0$.
For such a configuration,
an effective radius
$R_{\rm eff}\ge R_{\rm OV}$
may be defined so that
  $\Phi(R_{\rm eff})=0$.
 Although the density does not vanish at this point,
most of the mass will be contained inside the sphere
 of radius $R_{\rm eff}$.
 If we choose the boundary at
 $R\gg R_{\rm OV}$,
 the total mass will be dominated by the
 density distribution within $R_{\rm eff}$,
 and it will be almost independent of the
 choice of $R$.
 Thus,
 in the following we will work with
 $R=100\simeq 30 R_{\rm OV}$.

\begin{figure}
\vbox{\vskip230pt
\includegraphics{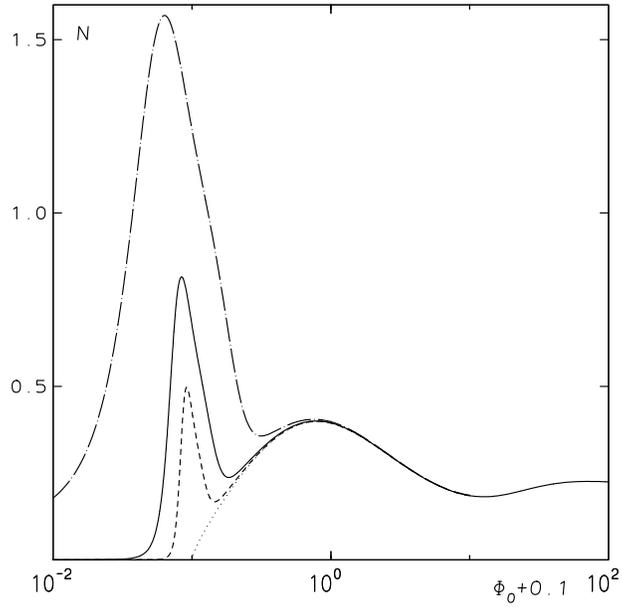}}
\caption{Fermion number $N$versus central potential $\Phi_0$
for $\alpha=\frac{\mu}{T}=300$ (full line)
500 (dashed line)
and 150 (dot-dashed line).
T=0 is represented by dotted line.}
\label{fig2}
\end{figure}

In Fig.~\ref{fig2}, the fermion number $N$ is plotted as a function of
initial $\Phi_0$ for several values of the parameter
$\alpha$.
In contrast to the $T=0$ case, all curves with
finite $\alpha$ have a peak around
$\Phi=0$.
The second peak corresponds to the OV
limit.
From this figure we can deduce that, for a given
$N$, there is a range of $\alpha$'s
for which the Thomas-Fermi equation may have
more than one solution.
This is a clear indication for
the existence of an
instability even below the
OV limit
and, as a consequence, we expect that
 a first-order phase transition occurs.

\begin{figure}
\vbox{\vskip230pt
\includegraphics{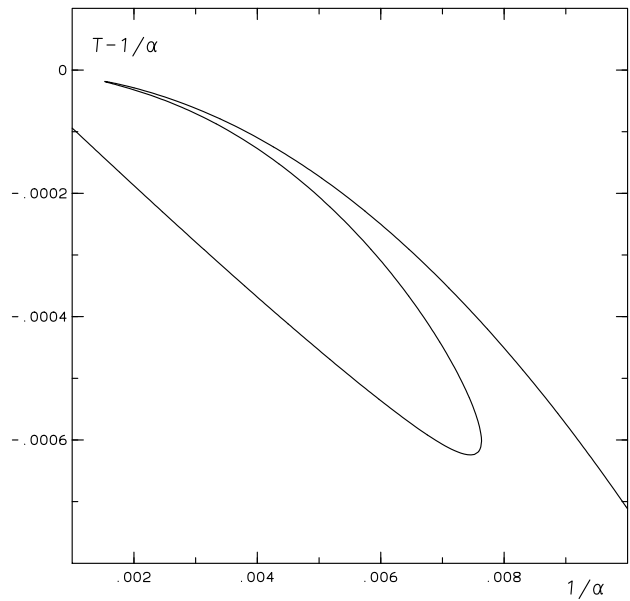}}
\caption{Temperature T (in units of $m$) versus $1/\alpha$
for $N=0.38$ and $R=100$.}
\label{fig3}
\end{figure}

\begin{figure}
\vbox{\vskip230pt
\includegraphics{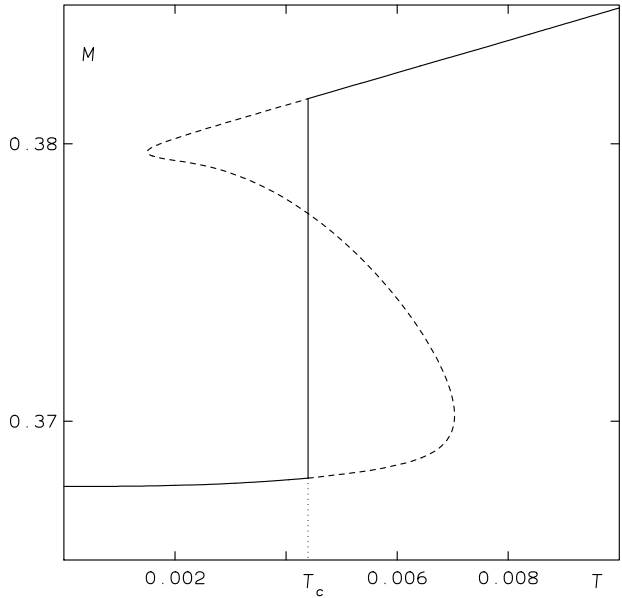}}
\caption{Total mass (in units of $b$) versus temperature
for $N=0.38$ and $R=100$.}
\label{fig4}
\end{figure}

\begin{figure}
\vbox{\vskip230pt
\includegraphics{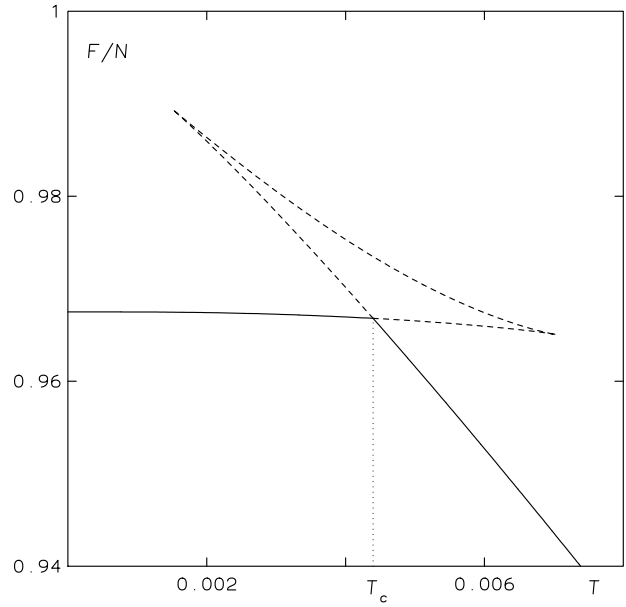}}
\caption{Free energy per fermion $F/N$
(in units of $m$)
 versus temperature T
for $N=0.38$ and $R=100$.}
\label{fig5}
\end{figure}

\begin{figure}
\vbox{\vskip230pt
\includegraphics{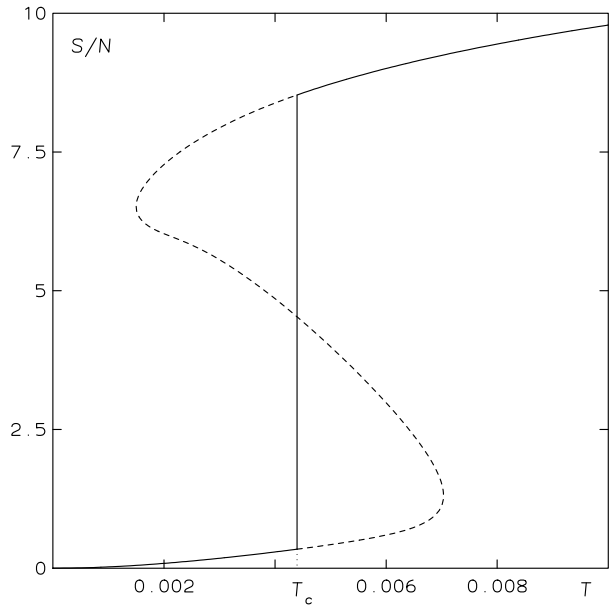}}
\caption{Entropy per fermion $S/N$ versus temperature $T$
for $N=0.38$ and $R=100$.}
\label{fig6}
\end{figure}

Fixing $N=0.38$, which is slightly below
the OV limit, we can now plot the temperature as a
function of $\alpha$ in Fig.~\ref{fig3}.
Using this figure as a parametric function for temperature,
 the mass, free energy, and entropy
are shown as
functions of temperature in Figs.
\ref{fig4},
\ref{fig5},
and
\ref{fig6}, respectively.
In the temperature interval
 $T=(0.0015-0.007) m$
there are
three distinct solutions of which
 only two are physical,
 namely, those
 for which the free energy
  assumes a minimum.
 The solution
 that can be continuously extended to any temperature
 above the mentioned interval
 is referred to
 as ``gas", whereas the solution that
  continues to exist at  low
 temperatures, and eventually becomes a degenerate
 Fermi gas at $T=0$, will be called ``condensate".
 In Fig.~\ref{fig2} the gas is represented
 on each curve by the
 part left from the first maximum,
 while
 the part from
 the first minimum up to the second maximum
 represents the condensate.
 By noting that $\Phi_0$ is negative
 for the gas and positive for the condensate,
 we may define an order parameter as
 \begin{equation}
 \delta=\Phi_0+|\Phi_0| ,
 \label{eq99}
 \end{equation}
 which is strictly positive in
 the condensed phase (ordered phase)
 and equal to zero in the gaseous phase
 (disordered phase).

The phase transition takes place at the
temperature $T_c$,
where the free energy of the gas and
that of the condensate
become equal.
The dashed curves in Figs.
\ref{fig4},
\ref{fig5},
and
\ref{fig6} represent the physically unstable solution.
In our example,
the transition temperature is
$T_c=0.0043951 m$,
as indicated in the plots by the dotted line.
The latent heat per fermion released
during the phase transition is given by the
mass difference at the point of
discontinuity
\begin{equation}
\frac{\Delta M}{N}=0.0438 m .
\end{equation}

So far, we have studied, as an example, an object with
number of fermions $N$ just below the OV limit
$N_{\rm OV}$.
 Any  object with $N<N_{\rm OV}$
will undergo a gravitational phase transition
at a
critical temperature which depends on the mass, of course.
With decreasing $N$,
the cavity  radius $R$
must be appropriately increased, since
the effective radius of the condensate increases
following approximately the
 zero-temperature
mass-radius relation.
As $N$ becomes smaller,
the system approaches
the nonrelativistic scaling regime
discussed in Sec.  \ref{sec2},
and for
$N\ll N_{\rm OV}$
the critical temperature will decrease according to
\begin{equation}
T_c={\rm const}\, N^{4/3},
\end{equation}
if the cavity  radius $R$ is simultaneously rescaled as $N^{-1/3}$.
In Fig.~{\ref{fig7}
we compare the critical temperature calculated
in both Newtonian and general-relativistic Thomas-Fermi models,
as a function of $N$.
The nonrelativistic scaling law turns out to be very accurate
for $N<0.2\,N_{\rm OV}$.
\begin{figure}
\vbox{\vskip230pt
\includegraphics{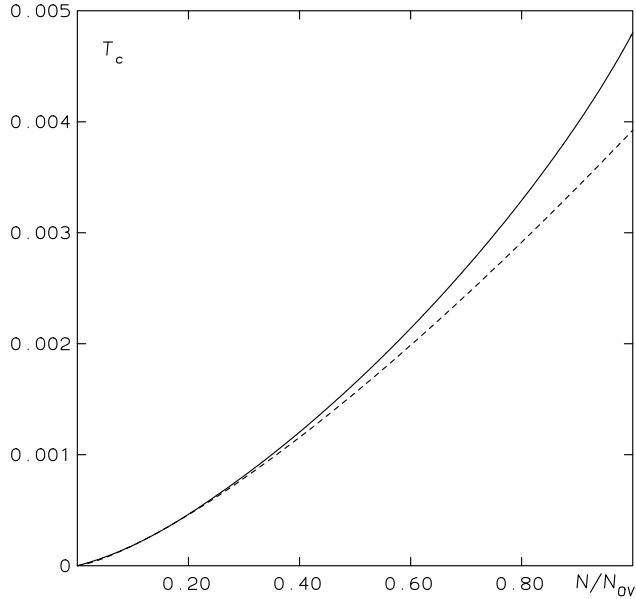}}
\caption{Critical temperature as a function
of the fermion number
in the Newtonian (dashed line) and general-relativistic
Thomas-Fermi model (solid line).}
\label{fig7}
\end{figure}

It is important to check that the critical
temperature is not very sensitive to
 variations of the cavity radius $R$
 for the following two reasons:
First, in our model $R$ is arbitrary except for the
requirement that it should be much larger then the
effective radius which
for
 $N=0.38$ is of the order
$R_{\rm eff}\simeq R_{\rm OV}=3.357$.
Second,
if the critical temperature rapidly  decreases with
$R$, the adiabatic cooling of the gas through the
universe expansion may not necessarily lead to
the point of the phase transition.
 Fig.~\ref{fig8} shows that the critical
temperature indeed decreases very slowly,
roughly by a factor of two if $R$ increases
from 30 to 300.
This is much weaker than the adiabatic cooling
of a nonrelativistic gas
which goes approximately as $1/R^2$.
Thus we conclude that the gravitational phase transition
will necessarily take place in the course of
the universe expansion.
\begin{figure}
\vbox{\vskip230pt
\includegraphics{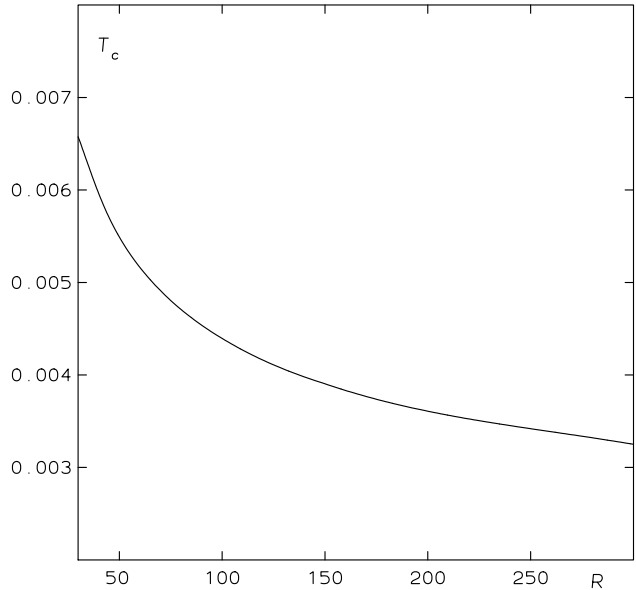}}
\caption{Critical temperature $T_c$ as a function
of the cavity radius $R$
for $N=0.38$.}
\label{fig8}
\end{figure}

\section{Conclusions}
\label{sec6}

In this work,
we have extended the Thomas-Fermi model
to a general-relativistic
 framework.
 This model was
 then applied
to  a system of self-gravitating
 fermions.
 We have
  investigated numerically
the circumstances under which
this system undergoes
a gravitational phase transition
 at nonzero temperature.
This phase transition is quite distinct from
the more extensively investigated strong-interaction driven phase
transition that might occur in neutron stars
\cite{saw,wit}.
The main underlying physics here
is the competition between the
partial degeneracy pressure due to the
Fermi-Dirac statistics and the
attractive force due to the gravitational interaction.
 It is obvious that the application of this model to
 astrophysical systems
 will work very well if the non-gravitational interactions
 between the individual particles can be neglected.
This is certainly the case, e.g., for
weakly interacting quasidegenerate heavy neutrino
neutralino, or gravitino matter \cite{bil2,bil1,cha,kul,bil3},
but perhaps it could be valid even for collisionless stellar systems
\cite{shu,chav}.

Finally, let us briefly comment on a
similar model by
Chau, Lake and Stone \cite{cha} which was considered
earlier
in the context of a possible galactic massive neutrino halo.
Their model differs  from our approach in essentially two aspects:
First, the equation of state is not consistent with
the condition of thermodynamical and chemical
equilibrium, i.e. with Eq.~(\ref{eq24a})
and second,
the particle number constraint (\ref{eq45})
is not imposed.
Thus, in contrast to the Thomas-Fermi model
discussed here,
the Chau et al.
 model does not describe a
canonical system in
equilibrium.

\subsection*{Acknowledgment}
We acknowledge useful discussions with
 D. Tsiklauri.
 This work  was supported by
 the Foundation for Fundamental Research
 (FFR).
\appendix

\section{Free Energy}
\label{app1}
Consider a canonical ensemble that describes
a nonrotating fluid
in equilibrium
at nonzero temperature.
We denote by
 $u_{\mu}$ ,
$p$, $\rho$, $n$, and $\sigma$ the velocity, pressure,
energy density, particle number density, and
entropy density of the fluid.
A canonical ensemble is subject to the
constraint
that the total number of particles
\begin{equation}
\int_{\Sigma} n\, u^{\mu}d\Sigma_{\mu}
=N
\label{eq26}
\end{equation}
should be fixed.
The spacelike hypersurface
$\Sigma$ that contains
the fluid is orthogonal to the time-translation
Killing vector field $k^{\mu}$
which is related to the velocity of the fluid
by
\begin{equation}
k^{\mu}=\xi u^{\mu}\, ;  \;\;\;\;\;\;
\xi=(k^{\mu}k_{\mu})^{1/2}.
\label{eq52}
\end{equation}
The energy-momentum tensor is defined as
\begin{equation}
T_{\mu\nu}=(p+\rho)u_{\mu}u_{\nu}-pg_{\mu\nu}.
\label{eq10}
\end{equation}
The metric generated by the mass distribution
in equilibrium
is  static, spherically symmetric, and asymptotically
flat, i.e.,
\begin{equation}
ds^2=\xi^2 dt^2 -\lambda^2 dr^2 -
     r^2(d\theta^2+\sin \theta d\phi^2).
\label{eq12}
\end{equation}
The metric coefficients
 $\xi$ and $\lambda$ may be determined from
Einstein's field equations.
However, we shall for the moment assume that
$\xi$ is an arbitrary function of $r$,
which asymptotically approaches unity,
and we parameterize $\lambda$ in terms of the mass
$\cal M$ within the radius $r$
\begin{equation}
\lambda=\left(1-\frac{2{\cal{M}}(r)}{r}\right)^{-1/2},
\label{eq16}
\end{equation}
where
\begin{equation}
{\cal{M}}(r)=\int^r_0 dr'\, 4\pi r'^2 \rho(r') \, .
\label{eq18}
\end{equation}

The temperature $\bar{T}$ and chemical potential $\bar{\mu}$
are, in general, metric-dependent local quantities.
The local law of energy-momentum conservation,
$T^{\mu\nu}_{\;\;\;\; ;\nu}=0$,
for a perfect fluid in a static gravitational field,
yields
the equation of hydrostatic equilibrium
\cite{lan}
\begin{equation}
\partial_{\nu}p=-(p+\rho)\xi^{-1}\partial_{\nu}\xi .
\label{eq20}
\end{equation}
This equation, together with the thermodynamic identity
(Gibbs-Duhem relation)
\begin{equation}
 d\frac{\bar{\mu}}{\bar{T}}=\frac{1}{n}
 d\frac{p}{\bar{T}}+\frac{\rho}{n}d\frac{1}{\bar{T}},
\label{eq21}
\end{equation}
and the condition that the heat flow
and diffusion vanish
\begin{equation}
 \frac{\bar{\mu}}{\bar{T}}={\rm const.} ,
\label{eq22}
\end{equation}
implies
\begin{equation}
\bar{T} \xi=T\, ; \;\;\;\;\;\;
\bar{\mu} \xi=\mu \, ,
\label{eq24}
\end{equation}
where $T$ and $\mu$ are constants equal to the
temperature and chemical potential at infinity.
The temperature $T$ may be chosen arbitrarily
as the temperature of the heatbath.
The quantity $\mu$ in a canonical ensemble is
an implicit functional of $\xi$
owing to the constraint
(\ref{eq26}).
The first equation in (\ref{eq24}) is the well known-Tolman condition
for thermal equilibrium in a gravitational
field \cite{tol}.

Following Gibbons and Hawking \cite{gib}, we postulate
that
the free energy of the canonical ensemble is
\begin{equation}
F=M-\int_{\Sigma} \bar{T}\sigma \, k^{\mu}d\Sigma_{\mu} \, ,
\label{eq50}
\end{equation}
where $M$ is the total mass as measured from infinity.
 The entropy density of a relativistic fluid may be expressed as
\begin{equation}
\sigma=\frac{1}{\bar{T}}(p+\rho-\bar{\mu} n),
\label{eq28}
\end{equation}
 $\bar{T}$ and
$\bar{\mu}$ being
the local temperature and chemical potential
as defined in (\ref{eq24}).
Based on Eq.~(\ref{eq24}), the free energy may be written
in the form analogous to ordinary thermodynamics
\begin{equation}
F=M-T S,
\label{eq54}
\end{equation}
with  $M={\cal{M}}(R)$,
and the total entropy $S$ defined as
\begin{equation}
S = \int_0^R dr\,4\pi r^2 \lambda
\frac{1}{\bar{T}}(p+\rho)-\frac{\mu}{T} N ,
\label{eq56}
\end{equation}
where we have used spherical symmetry to
replace the proper volume integral as
\begin{equation}
\int_{\Sigma} u^{\mu}d\Sigma_{\mu}
= \int_0^R dr 4\pi r^2 \lambda .
\label{eq58}
\end{equation}

The following theorem relates
the extrema of the free energy
with the solutions of Einstein's field equation:
\newpage

\noindent
{\bf Theorem:}
{\it
Among all momentarily
static, spherically symmetric configurations
$\{\xi(r),{\cal{M}}(r)\}$
which, for a given temperature $T$ at infinity,
 contain a specified number of particles
\begin{equation}
 \int_0^R 4\pi r^2 dr \, \lambda(r)  n(r) = N
\label{eq60}
\end{equation}
within a spherical volume of a given radius
 $R$,
those and only those  configurations
that
extremize the quantity F defined by
{\rm (\ref{eq54})}
 will
 satisfy Einstein's field equation
\begin{equation}
\label{eq62}
\frac{d\xi}{dr}=
\xi\frac{{\cal M}+4\pi r^3 p}{r(r-2{\cal{M}})} \, ,
\end{equation}
with the boundary condition
\begin{equation}
\xi(R)=\left(1-\frac{2 M}{R}\right)^{1/2}.
\label{eq63}
\end{equation}
}

\noindent
{\bf Proof:}
By making use of the identity (\ref{eq21})
 and the fact
that $\delta(\bar{\mu}/\bar{T})=\delta(\mu/T)$
and that $N$ is fixed by the constraint (\ref{eq60}),
from Eqs.~(\ref{eq54}) and (\ref{eq56})
we find
\begin{equation}
\delta F= \delta M -
\int_0^R dr\, 4\pi r^2 \frac{T}{\bar{T}}(p+\rho)
\delta \lambda
-  \int_0^R dr\, 4\pi r^2 \lambda \frac{T}{\bar{T}} \delta\rho \, .
\label{eq64}
\end{equation}
The variations $\delta\lambda$ and $\delta\rho$
 can be expressed in terms of the variation
$\delta {\cal{M}}(r)$
and its derivative
\begin{equation}
\frac{d\delta {\cal M}}{dr} =4\pi r^2 \delta\rho,
\label{eq66}
\end{equation}
yielding
\begin{equation}
\delta F= \delta M -
\int_0^R dr\, 4\pi r^2
 \frac{T}{\bar{T}}(p+\rho)
\frac{\partial\lambda}{\partial{\cal M}}
\delta{\cal M}
-\int_0^R dr\, \lambda\frac{T}{\bar{T}}\frac{
\delta{\cal M}}{dr} \, .
\label{eq68}
\end{equation}
By partial integration of the last term,
and replacing $T/\bar{T}$ by $\xi$, we find
\begin{equation}
\delta F =
\left[1-\lambda(R)\xi(R)\right]\delta M
- \int_0^R dr\, \left[4\pi r^2 \xi (p+\rho)
\frac{\partial\lambda}{\partial{\cal M}}
-\frac{d}{dr}(\lambda\xi)\right]\delta{\cal M} \, ,
\label{eq70}
\end{equation}
where $\delta{\cal{M}}(r)$ is an arbitrary variation
on the interval $[0,R]$,
except for
the constraint
$\delta{\cal{M}}(0)=0$.
Therefore, $\delta F$ will vanish if and only if
\begin{equation}
4\pi r^2 \xi (p+\rho)
\frac{\partial\lambda}{\partial{\cal M}}
-\frac{d}{dr}(\lambda\xi) =0
\label{eq72}
\end{equation}
and
\begin{equation}
1-\lambda(R)\xi(R) =0.
\label{eq74}
\end{equation}
Using Eqs.~(\ref{eq16}) and (\ref{eq18}), we can write
Eq.~(\ref{eq72}) in the form (\ref{eq62}),
and Eq.~(\ref{eq74}) gives the desired boundary condition
 (\ref{eq63}).
Thus,
$\delta F=0$ if and only if a configuration
$\{\xi,{\cal M}\}$ satisfies Eq.~(\ref{eq62})
with (\ref{eq63}),
as was to be shown.
\vskip 0.1in

\noindent
{\it Remark 1:}
A solution  to Eq.~(\ref{eq62})
is dynamically stable if
the free energy assumes a minimum.
\\
{\it Remark 2:}
Our Theorem 1 is a finite-temperature generalization
of the result obtained for
cold, catalyzed matter \cite{har}.

%
%

%

%
\end{document}